\newif\ifeprint \eprinttrue                                          %%
\PassOptionsToPackage{obeyspaces,spaces,hyphens}{url}
\ifeprint                                                            %%
\documentclass[rmp,twocolumn,floatfix,                               %%
  showkeys,letterpaper,amsmath]{revtex4}                             %%
\usepackage{hyperref,orcidlink}                                      %%
\hypersetup{pdftitle={The area of rhumb polygons},                   %%
  pdfauthor={C. F. F. Karney}}                                       %%
\else                                                                %%
\documentclass[pdflatex]{NAV}
\renewcommand{\S}{\textsection}

\fi                                                                  %%
\usepackage{array,graphicx,amssymb}
\newcommand{\doi}[1]{\href
  {https://doi.org/#1}{doi:\discretionary{}{}{}#1}}
\renewcommand{\d}{\mathrm d}
\newcommand{\gd}{\mathop{\mathrm{gd}}\nolimits}
\newcommand{\lam}{\mathop{\mathrm{lam}}\nolimits}
\newcommand{\atanx}[2]{\tan^{-1}\genfrac{}{}{1.4pt}{0}{#1}{#2}}
\newcommand{\divdiff}{\mathord{\ooalign{$\triangle$\cr\hfil$\mid$\hfil\cr}}}

\usepackage{breakurl}
\ifpdf\def\urlalt#1#2{\burlalt{#2}{#1}}\else\let\urlalt=\burlalt\fi
\newcommand{\dlmf}[2]{\urlalt{http://dlmf.nist.gov/#2}{#1}}
\def\figuredir{figures}
\begin{document}
\title[The area of rhumb polygons]{The area of rhumb polygons}
\ifeprint                                                            %%
\author{Charles F. F. Karney\,\orcidlink                             %%
  {0000-0002-5006-5836}}                                             %%
\email{charles.karney@sri.com}                                       %%
\affiliation{SRI International,                                      %%
  Princeton, NJ 08540-6449, USA}                                     %%
\date{\today}                                                        %%
\else                                                                %%
\author[Charles F. F. Karney]{Charles F. F. Karney$^1$}
\email{charles.karney@sri.com}
\address{\add{1}{{SRI International,
  Princeton, NJ 08540-6449, USA}}}
\fi                                                                  %%

\begin{abstract}
The formula for the area of a rhumb polygon, a polygon whose edges are
rhumb lines on an ellipsoid of revolution, is derived and a method is
given for computing the area accurately.  This paper also points out
that standard methods for computing rhumb lines give inaccurate results
for nearly east- or west-going lines; this problem is remedied by the
systematic use of divided differences.
\end{abstract}

\maketitle

\section{Introduction}

\begin{figure}[tb]
\begin{center}% produced by rhumblines.m
\ifeprint\mbox{}\\[1ex]\fi      %%
\includegraphics[scale=0.75]{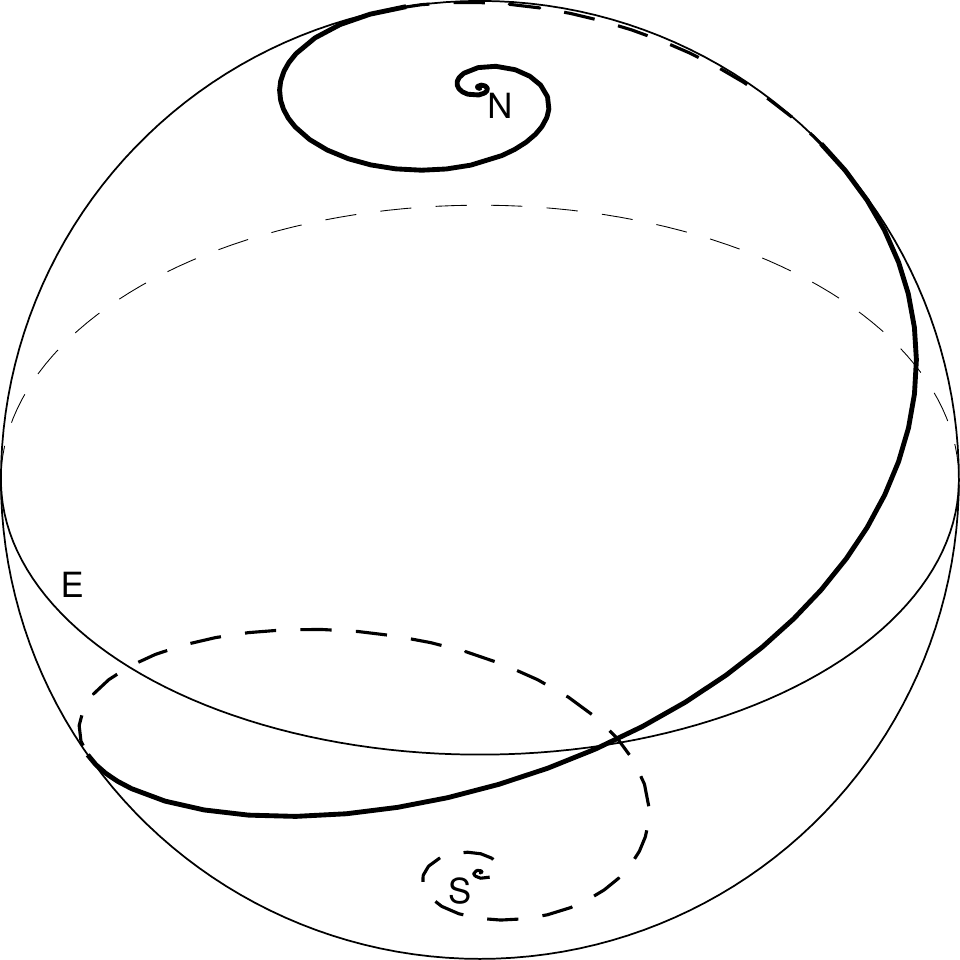}
\end{center}
\caption{\label{rhumbline-sphere}
A rhumb line on a sphere illustrating its distinctive characteristic:
spiraling infinitely many times around each pole (in the polar
stereographic projection, the rhumb line maps to a logarithmic spiral).
However, the rhumb line distance between the poles is finite; in this
case, the constant azimuth is taken to be $\sec^{-1}3 = 70.5^\circ$ so
that this distance is 3 times the distance along the meridian.  The
rhumb line is viewed in an orthographic projection from a point at
latitude $35^\circ$; the poles and the equator are labeled ``N'', ``S'',
and ``E''.}
\end{figure}

\begin{figure}[tb]
\begin{center}% produced by rhumblines.m
\includegraphics[scale=0.75]{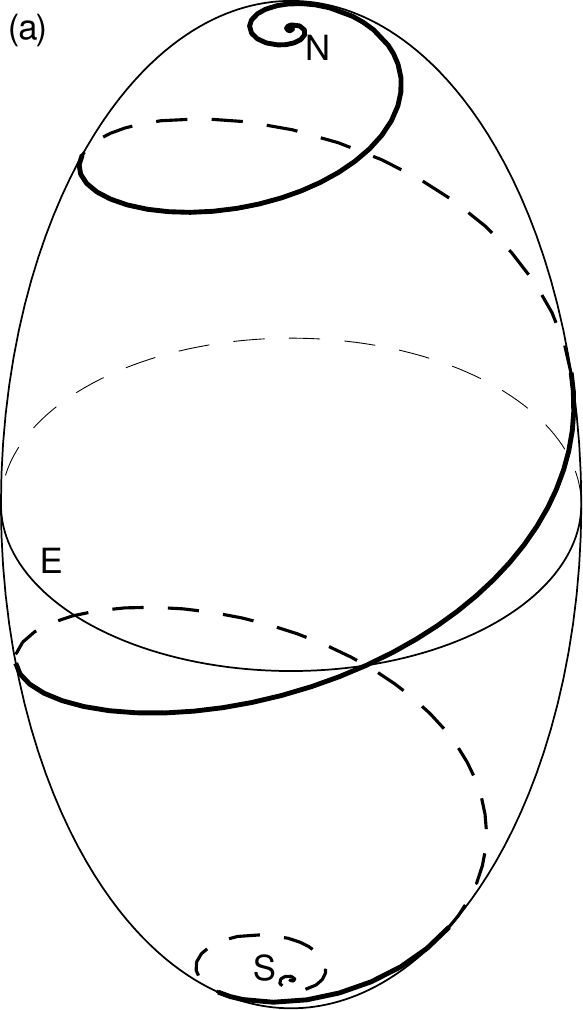}\\[5ex]
\includegraphics[scale=0.75]{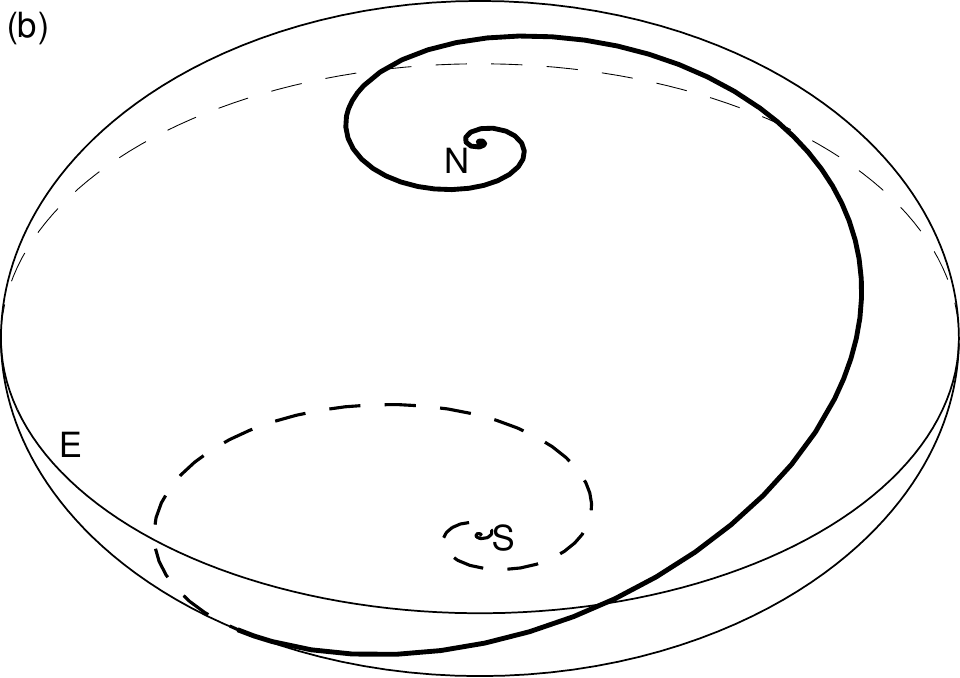}
\end{center}
\caption{\label{rhumblines-ellipsoid}
Rhumb lines (a) on a prolate ellipsoid, $b/a = 2$, and (b) on an oblate
ellipsoid, $b/a = \frac12$.  The bearing of the rhumb line and the
projection match that of Fig.~\ref{rhumbline-sphere}.}
\end{figure}

A rhumb line (also called a loxodrome) is a line of constant azimuth on
the surface of an ellipsoid, see Figs.~\ref{rhumbline-sphere}
and \ref{rhumblines-ellipsoid}.  It is important for historical reasons
because sailing with a constant compass heading is simpler than sailing
the shorter geodesic course; the distinguishing feature of the Mercator
projection is that rhumb lines map to straight lines in this projection.
The method for computing rhumb lines on a sphere was given by Wright in
1599 and this was extended to an ellipsoid of revolution
by \citet{lambert72} who derived the ellipsoidal generalization of the
Mercator projection.  Distances along the rhumb line are then given in
terms of the incomplete elliptic integral of the second
kind \citep{legendre11}.  Subsequent papers on rhumb lines largely focus
on simplifications to allow them to be easily computed by navigators at
sea.

In this paper we address two issues:
\begin{enumerate}
\item
Computing rhumb lines accurately for the case of headings close to
$\pm90^\circ$.  The standard method results in a severe loss of
significant bits.  This can be avoided by the use of divided
differences.
\item
Finding the area ``under'' a rhumb line segment, i.e., the area of the
quadrilateral formed by the segment, two meridian arcs, and a segment of
the equator.  This allows the areas of rhumb polygons to be computed
accurately even for very eccentric ellipsoids.
\end{enumerate}
We consider an ellipsoid of revolution with equatorial radius $a$ and
polar semi-axis $b$.  The shape of the ellipsoid is variously
characterized by the flattening $f = (a-b)/a$, the third flattening $n =
(a-b)/(a+b)$, the eccentricity squared $e^2 = (a^2-b^2)/a^2$, or the
second eccentricity squared $e'^2 = (a^2-b^2)/b^2$.  The (geographic)
latitude and longitude are denoted by $\phi$ and $\lambda$.  The azimuth
of a rhumb line, measured clockwise from north is denoted by $\alpha$.

\section{Formulation of the rhumb line problem} \label{form}

Rhumb lines map to straight lines in the Mercator projection because
this projection is conformal and because it maps meridians to parallel
straight lines.  Ellipsoidal effects can be simply introduced using
various auxiliary latitudes, the conformal latitude $\chi$, the
rectifying latitude $\mu$, the parametric latitude $\beta$, and the
authalic latitude $\xi$.  In the spherical limit, these four auxiliary
latitudes reduce to the geographic latitude $\phi$.

It is possible to convert between these various latitudes
accurately \citep{karney-auxlat} and, in this paper, we treat this as a
solved problem.  We keep the notation simple by writing, for example,
$\mu_1$ instead of $\mu(\phi_1)$.

The Mercator projection maps a point on the earth with longitude
$\lambda$ and latitude $\phi$ to a point $(\lambda, \psi)$ on the plane
where $\psi$ is the isometric latitude
\begin{equation}\label{psi-chi}
\psi = \lam\chi = \gd^{-1}\chi = \sinh^{-1} \tan \chi,
\end{equation}
$\lam\chi$ is the Lambertian function, and its inverse, $\gd \psi
= \tan^{-1} \sinh \psi$, is the gudermannian function
\citep[\S\dlmf{4.23(viii)}{4.23.viii}]{dlmf10}.

Treating first the inverse rhumb line problem, finding the course
$\alpha_{12}$ and length $s_{12}$ of the rhumb line between two points
$(\phi_1, \lambda_1)$ and $(\phi_2, \lambda_2)$, we have simply
\begin{align}
\alpha_{12} &= \atanx{\lambda_{12}}{\psi_{12}}, \label{azi12}
\displaybreak[0]\\
\frac{s_{12}}R &= \mu_{12} \sec\alpha_{12} \notag\\
&= \frac{\mu_{12}}{\psi_{12}}\sqrt{\lambda_{12}^2 + \psi_{12}^2}\label{s12},
\end{align}
where $R$ is the rectifying radius of the ellipsoid, $\psi_{12} = \psi_2
- \psi_1$, $\mu_{12} = \mu_2 - \mu_1$, $\lambda_{12} = \lambda_2
- \lambda_1$; typically, $\lambda_{12}$ is reduced to the range
$[-\pi,\pi]$ (thus giving the course of the {\it shortest} rhumb line).
The heavy ratio line in Eq.~(\ref{azi12}) indicates that the quadrant of
the arctangent is determined by the signs of the numerator and
denominator of the ratio.  The distance from the equator to a pole can
be expressed in terms of $R$ as
\begin{equation}
\frac\pi2 R = aE(e) = bE(ie') = 2 R_G(0,a^2,b^2),
\end{equation}
where $E(k)$ is the complete elliptic integral of the second
kind \citep[\S\dlmf{19.2(ii)}{19.2.ii}]{dlmf10} and $R_G(x,y,z)$ is one
of the symmetric elliptic
integrals \citep[\S\dlmf{19.16(ii)}{19.16.ii}]{dlmf10}.  For small
flattening, we can use \citep{ivory98}
\begin{align}
R &= \frac a{1+n}
  \sum_{j=0}^\infty \biggl(\frac{(2j-3)!!}{(2j)!!}\biggr)^2 n^{2j}
  \notag\\
  &= \frac a{1+n} \bigl(\textstyle 1+\frac14 n^2 + \frac1{64} n^4 +
  \frac1{256} n^6 + O(n^8)\bigr).
\end{align}

The direct problem is treated similarly.  We are given the
starting point $(\phi_1, \lambda_1)$, the course $\alpha_{12}$, and the
distance $s_{12}$ along a rhumb line, and wish to determine the
destination $(\phi_2, \lambda_2)$.  We can write
\begin{equation}
\mu_{12} = \frac{s_{12}}R\cos\alpha_{12},
\end{equation}
from which we can find $\mu_2$, $\phi_2$ and $\psi_2$.  Then we have
\begin{align}
\lambda_{12} &= \psi_{12} \tan\alpha_{12} \notag\\
&= \frac{s_{12}}R \frac{\psi_{12}}{\mu_{12}}\sin\alpha_{12}.\label{lam12}
\end{align}
In this formulation, we do not restrict the sign of $s_{12}$; negative
distances are allowed in the direct problem.  If the value of $\mu_2
= \mu_1 + \mu_{12}$ lies outside the normal range for latitudes,
$[-\frac12\pi,\frac12\pi]$, then it should be replaced by its
supplement.  Because, in this case, the rhumb line has spiraled
infinitely many times around the pole, see Figs.~\ref{rhumbline-sphere}
and \ref{rhumblines-ellipsoid}, $\lambda_{12}$ is then indeterminate.
This is also the outcome if $\mu_2 = \pm\frac12\pi$ because then
$\psi_{12} = \pm\infty$ and Eq.~(\ref{lam12}) gives an indeterminate
result.

Both Eqs.~(\ref{s12}) and (\ref{lam12}) are indeterminate in the limit
$\phi_2 \rightarrow \phi_1$, since both $\psi_{12}$ and $\mu_{12}$
vanish.  However, we can apply l'H\^opital's rule and write
\begin{equation}\label{parallel}
\frac{\mu_{12}}{\psi_{12}} \rightarrow \frac{\d\mu_1}{\d\psi_1}
= \frac{a\cos\beta_1}R,
\end{equation}
where the formula for the derivative follows from the defining
differential equations for $\psi$ and $\mu$.  Note that $a\cos\beta_1$
is the radius of a circle of latitude $\phi_1$.

This provides a complete solution for the course of rhumb lines.  We can
see that by using auxiliary latitudes, the spherical solution is readily
generalized to an ellipsoid; Fig.~\ref{rhumblines-ellipsoid} shows rhumb
lines on prolate and oblate ellipsoids.  In implementing such a
solution, we have a choice of using series expansions for converting
between auxiliary latitudes, recommended for $\lvert
f \rvert \lesssim \frac1{100}$, or directly evaluating the formulas for
the auxiliary latitudes.  Details of both methods are given
in \citet{karney-auxlat}.

When using Eqs.~(\ref{s12}) and (\ref{lam12}), we are faced with a
severe loss of accuracy if $\phi_2 \approx \phi_1$ because then both
$\mu_{12}$ and $\psi_{12}$ involve the difference between two nearly
equal quantities resulting in catastrophic roundoff errors.

A simple solution, suggested, for example, by \citet{meyer11}, is to
extend the treatment of rhumb lines that run exactly along parallels to
the case $\phi_2\approx\phi_1$, i.e., to replace the ratio
$\mu_{12}/\psi_{12}$ by the derivative $\d\mu/\d\psi$,
Eq.~(\ref{parallel}), evaluated at the midpoint $\phi
= {\frac12(\phi_1+\phi_2)}$.  However this is not entirely satisfactory:
you have to pick the transition point where the derivative takes over
from the ratio of differences; and, near this transition point, many
bits of accuracy will be lost.

\begin{figure}[tb]
\begin{center}% produced by roundoff.m
\includegraphics[scale=0.75]{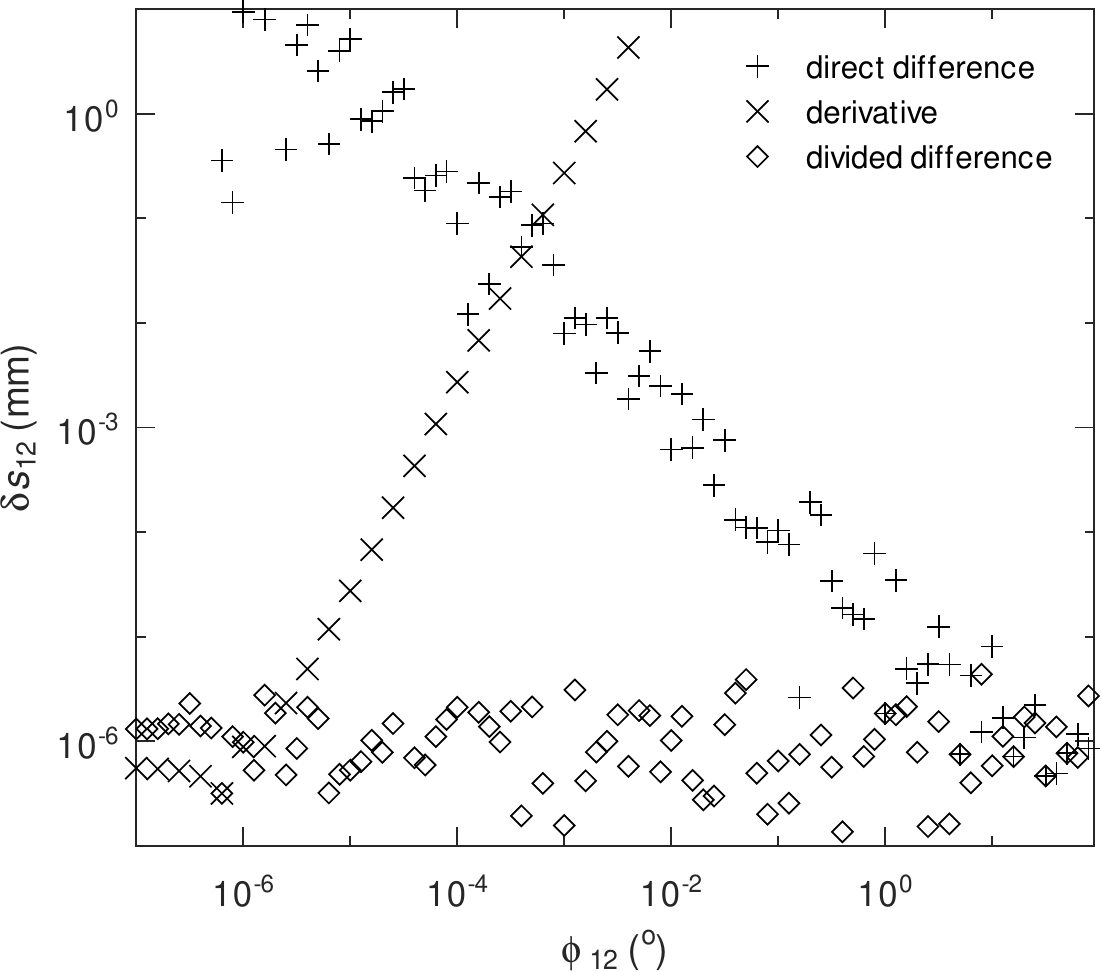}
\end{center}
\caption{\label{roundoff}
The absolute errors $\delta s_{12}$ in computing $s_{12}$ for rhumb
lines on a sphere of radius $a = 6400\,\mathrm{km}$, with mean
latitude $\frac12(\phi_1 + \phi_2) = 45^\circ$, longitude difference
$\lambda_{12} = 90^\circ$, and various values of $\phi_{12} = \phi_2
- \phi_1$.  Three methods of computing $\mu_{12}/\psi_{12}$ are
compared: the direct ratio of the differences Eq.~(\ref{s12}) (plus
signs); approximating the ratio by the derivative Eq.~(\ref{parallel})
(crosses); evaluating the ratio using divided differences
Eq.~(\ref{s12-divdiff}) (diamonds).  The systematic variation of the
errors for the derivative approximation is indicative that this is (for
the most part) a truncation error.  The random variations for the other
cases are due to roundoff.  In all three cases, the calculations
are carried out in double precision (53 bits in the fraction of the
floating-point representation); these are compared with the results of a
high (256-bit) precision calculation which are taken to be ``exact''.}
\end{figure}%
This conundrum is nicely illustrated by Fig.~\ref{roundoff} which shows
the errors $\delta s_{12}$ in the computed distance as $\phi_{12}
= \phi_2 - \phi_1$ is varied.  Here the plus signs and crosses give
$\delta s_{12}$ when computing $\mu_{12}/\psi_{12}$ using the ratio of
the differences and the mid-point derivative, respectively.  Given just
these two choices for the calculation, there is an unavoidable
degradation in the accuracy for $0.000\,01^\circ < \phi_{12} < 1^\circ$.
Selecting $\phi_{12} = 0.001^\circ$ as the transition point, the maximum
error in the distance is about $0.1\,\mathrm{mm}$.  This loss of
accuracy may not matter in some applications; however, we would wish to
avoid such a compromise if possible.

It turns out that we can do substantially better and maintain full
double-precision accuracy.  Indeed \citet{botnev14} provide formulas
which allow $\mu_{12}/\psi_{12}$ to be computed accurately; however
their treatment of $\psi_{12}$ is somewhat ad hoc and their formula for
$\mu_{12}$ applies only for small flattening.  The problem can be
addressed more simply by using the algebra of {\it divided
differences} \citep{kahan99}.  This technique can be used to reduce
roundoff errors in a wide range of applications and we detail how it can
be applied to computing $\mu_{12}/\psi_{12}$ in Appendix \ref{divided}.
The result is illustrated by the diamonds in Fig.~\ref{roundoff} which
show that the errors using divided differences with double precision are
on the order of $1\,\mathrm{nm}$ throughout the range of $\phi_{12}$.
This establishes that the conventional approach for computing the
distance (the plus signs and the crosses) needlessly loses, in some
cases, just over a quarter of the bits of precision in a floating-point
number.  Using divided differences allows for a single computational
method, avoids the tricky business of picking transition points, and
maintains full accuracy.

\section{The area under a rhumb line} \label{area}

The boundaries of political entities are sometimes defined as oblique
``straight'' lines, e.g., the border between California and Nevada, the
border between Algeria and its southern neighbors, Mauritania and Mali,
and the zigzag border between Jordan and Saudi Arabia.  It's often not
clear precisely how these lines are defined.  Reasonable choices are
geodesics (straight on the ellipsoid), rhumb lines (straight in the
Mercator projection), or straight lines in some other agreed-upon
projection.  Borders can also be specified as a circle of latitude
(which is, of course, also a rhumb line), for example, the 49th parallel
which separates the USA and Canada.  Assuming that the border of an
entity is defined in terms of rhumb lines, we might wish to compute the
resulting area.

This exercise was carried out for geodesic polygons in \citet{karney13},
where, following \citet{danielsen89}, the area between a geodesic line
segment and the equator was determined.  The area of a polygon is then
given by summing algebraically the contribution from each of the polygon
edges (with an adjustment required for polygons that encircle a pole).

Here, we repeat this exercise by computing $S_{12}$, the area of a rhumb
quadrilateral with vertices $(0, \lambda_1)$, $(0, \lambda_2)$,
$(\phi_2, \lambda_2)$, and $(\phi_1, \lambda_1)$ made up of two meridian
arcs, an equatorial segment, and a general rhumb segment.  With this
result in hand, we can compute the area of arbitrary rhumb polygons.
The area for a particular segment is given by
\begin{equation}
S_{12} = \int_{\lambda_1}^{\lambda_2} c^2 \sin\xi \,\d\lambda,\label{S-eq}
\end{equation}
where $\xi$ is the authalic latitude and
\begin{equation}
c^2 = \frac{a^2}2 + \frac{b^2}2 \frac{\tanh^{-1}e}e
\end{equation}
is the authalic radius squared.  We change the variable of integration
in Eq.~(\ref{S-eq}) first from $\lambda$ to $\psi$, using
$\d\lambda/\d\psi = \lambda_{12}/\psi_{12}$, and then to $\chi$, using
$\d\psi/\d\chi = \sec\chi$, to obtain
\begin{equation}\label{mean-xi}
S_{12} = c^2\lambda_{12}\frac{p_{12}}{\psi_{12}},
\end{equation}
where $p_{12} = p(\chi_2) - p(\chi_1)$ and
\begin{equation}\label{Q-int}
p(\chi) = \int \sin\xi\sec\chi \,\d\chi.
\end{equation}
In the limit, $\chi_2 \rightarrow \chi_1$, we have (applying
l'H\^opital's rule again)
\begin{equation}\label{lhopital2}
S_{12} \rightarrow c^2\lambda_{12}
\frac{\d p(\chi_1)/\d \chi_1}{\d \psi_1/\d \chi_1}
= c^2\lambda_{12}\sin\xi_1,
\end{equation}
the well-known result for the area under a segment of a
circle of latitude.

In the spherical limit, we have $\sin\xi \rightarrow \sin\chi$ and the
integrand in Eq.~(\ref{Q-int}) becomes $\tan\chi$.  This suggests that
we separate $p(\chi)$ into a spherical contribution and an ellipsoidal
correction by writing
\begin{align}\label{Q-eq}
p(\chi) &= p_0(\chi) + p_\chi(\chi),\displaybreak[0]\\
p_0(\chi) &= \int\tan\chi\,\d\chi \notag\\
&=\log\sec\chi = \sinh^{-1}\frac{\tan\chi\sin\chi}2, \label{p0-eq}
\displaybreak[0]\\
p_\chi(\chi) &= \int q_\chi(\chi) \,\d\chi, \label{DeltaQ}
\displaybreak[0]\\
q_\chi(\chi) &= \frac{\sin\xi-\sin\chi}{\cos\chi}. \label{q-eq}
\end{align}
The spherical term $p_0(\chi)$ is singular for
$\chi \rightarrow \frac12 \pi$; this just reflects the rapid encircling
of the pole by a rhumb line as it approaches a pole.  The second
expression for the integral in Eq.~(\ref{p0-eq}) avoids roundoff errors
for small~$\chi$.

\begin{figure}[tb]
\begin{center}% produced by qintegrand.m
\includegraphics[scale=0.75]{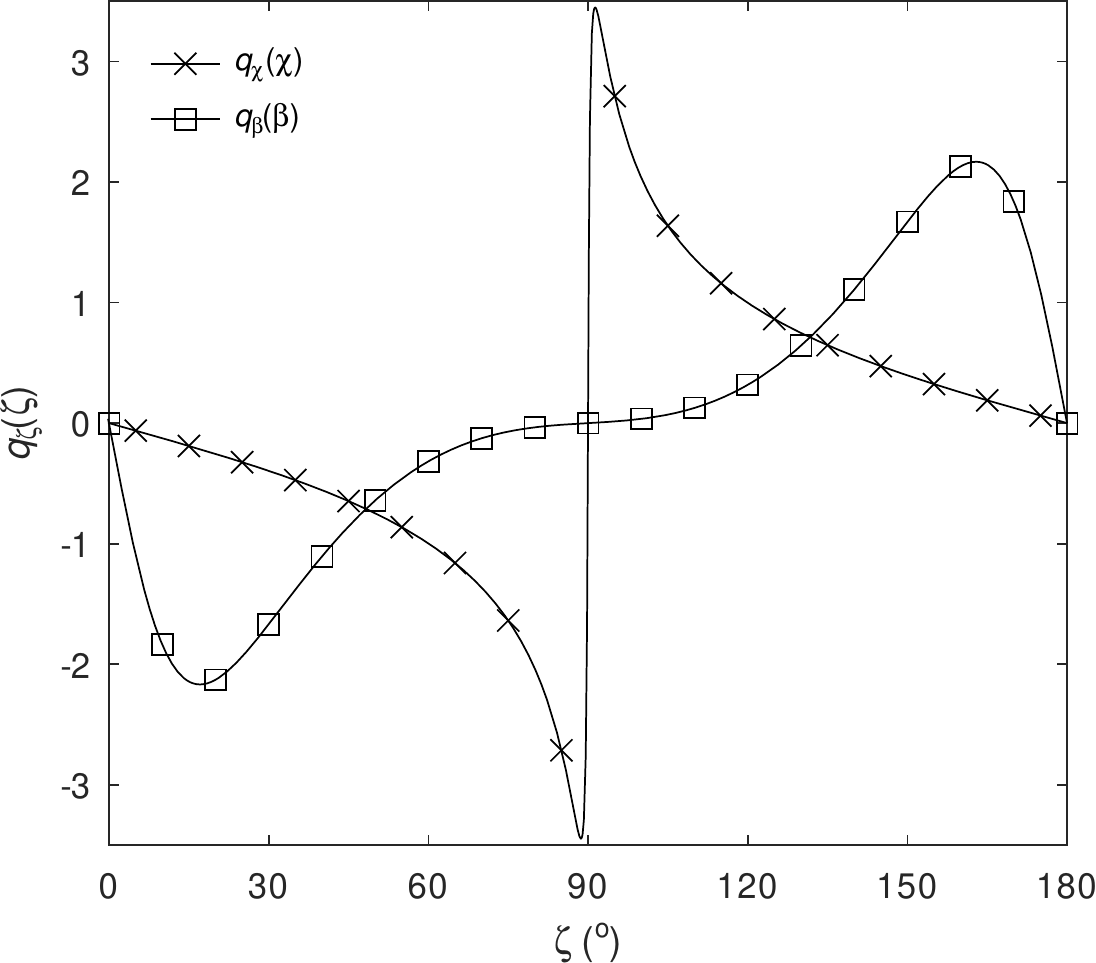}
\end{center}
\caption{\label{qintegrand}
The integrands $q_\chi(\chi)$ and $q_\beta(\beta)$, Eqs.~(\ref{q-eq})
and (\ref{q-eq-beta}), for the ellipsoidal correction for the area
integral marked, respectively, with crosses and squares.  The case
illustrated is for a prolate ellipsoid with $n = -\frac58$.  Note that
$\int_0^{\pi/2} q_\chi(\chi)\,\d\chi =
\int_0^{\pi/2} q_\beta(\beta)\,\d\beta$.}
\end{figure}%
We are now confronted with the evaluation of $p_\chi(\chi)$,
Eq.~(\ref{DeltaQ}).  Almost certainly the integral cannot be carried out
in closed form.  However, since $q_\chi(\chi)$ is a periodic function
(with period $\pi$), it is natural to develop it as a Fourier series
allowing the integral to be simply performed.  Unfortunately, even
though the integrand $q_\chi(\chi)$, Eq.~(\ref{q-eq}), is a well-behaved
function of $\chi$, it can vary strongly particularly for prolate
ellipsoids; for example, the curve marked with crosses in
Fig.~\ref{qintegrand} shows $q_\chi(\chi)$ for $n = -\frac58$ or $b/a =
13/3$.  As a result, the Fourier series converges slowly with many terms
needed to yield an accurate result.

We address this problem by changing the variable of integration in
Eq.~(\ref{DeltaQ}) from $\chi$ to the parametric latitude $\beta$.  Thus
we replace Eq.~(\ref{DeltaQ}) by
\begin{equation} \label{DeltaQ-beta}
p_\beta(\beta) = \int q_\beta(\beta) \,\d\beta,
\end{equation}
with
\begin{equation}\label{q-eq-beta}
q_\beta(\beta) = q_\chi(\chi) \frac{\d\chi}{\d\beta}
= (1-f) \frac{\sin\xi-\sin\chi}{\cos\phi},
\end{equation}
where we have used
\begin{equation}
\frac{\d\chi}{\d\beta} = (1-f)\frac{\cos\chi}{\cos\phi}.
\end{equation}
Our notation here is such that $p_\beta(\beta) =
p_\beta\bigl(\beta(\chi)\bigr) = p_\chi(\chi)$.  The curve marked with
squares in Fig.~\ref{qintegrand} shows $q_\beta(\beta)$ for $n =
-\frac58$ confirming that it varies more smoothly than $q_\chi(\chi)$.
Finally, we write
\begin{equation}\label{DeltaQ2}
p_\beta(\beta) = \sum_{l=1}^L P_l \cos 2l\beta,
\end{equation}
where $P_l$ are the coefficients of the Fourier series for
$p_\beta(\beta)$ and $L$ is picked sufficiently large that we may
approximate $P_l = 0$ for $l > L$.  For the case illustrated in
Fig.~\ref{qintegrand}, $n = -\frac58$, taking $L=50$ allows
$q_\beta(\beta)$ to be evaluated with full double-precision accuracy; in
contrast, we would have needed to include $L=L_\chi=3993$ terms if we
were fitting $q_\chi(\chi)$.  Since $p_\beta(\beta)$ is an indefinite
integral, Eq.~(\ref{DeltaQ-beta}), we pick the constant of integration
such that $P_0 = 0$ and the Fourier series, Eq.~(\ref{DeltaQ2}), starts
with the $l=1$ term.

If the flattening of the ellipsoid is small, we can expand
$q_\beta(\beta)$ as a Taylor series in $n$; this allows the integral to
be performed analytically, giving the following series approximations
for $P_l$:
\begin{equation}\label{QdotN}
\mathbf P^{(L)} =
  \mathsf Q^{(L)} \cdot \mathbf N^{(L)} + O(n^{L+1}),
\end{equation}
where
\begin{align}
\mathbf P^{(L)} &= [P_1, P_2, P_3, \ldots, P_L]^{\mathsf T},\\
\mathbf N^{(L)} &= [n, n^2, n^3, \ldots, n^L]^{\mathsf T}
\end{align}
are column vectors of length $L$ and, for $L=6$,
\begin{equation}\label{Qeq}
\renewcommand{\arraystretch}{1.3}
\mathsf Q^{(6)} =
\begin{bmatrix}
-\frac{1}{3}&\frac{22}{45}&-\frac{398}{945}&\frac{596}{2025}&-\frac{102614}{467775}&\frac{138734126}{638512875}\\
&\frac{1}{5}&-\frac{118}{315}&\frac{1543}{4725}&-\frac{24562}{155925}&\frac{17749373}{425675250}\\
&&-\frac{17}{315}&\frac{152}{945}&-\frac{38068}{155925}&\frac{1882432}{8513505}\\
&&&\frac{5}{252}&-\frac{752}{10395}&\frac{268864}{2027025}\\
&&&&-\frac{101}{17325}&\frac{62464}{2027025}\\
&&&&&\frac{11537}{4054050}
\end{bmatrix}
\end{equation}
is an $L\times L$ upper triangular matrix.  (Note that $P_l = O(n^l)$,
so that all the neglected, $l>L$, terms in the Fourier series for
$p_\beta(\beta)$ are $O(n^{L+1})$.)  Here the necessary algebra for
carrying out the Taylor expansion and subsequent integration was
performed by \citet{maxima}.  Truncating the expansion at $L=6$ as shown
here gives full double-precision accuracy for $\lvert
f\rvert \le \frac1{100}$.

The radius of convergence of this series can be estimated by examining
the coefficients $\mathsf Q^{(40)}$, which indicates convergence for
$\lvert n\vert \lesssim 1$, i.e., for all eccentricities.  Nevertheless,
it is wise to limit the series method to reasonably small values of
the flattening, say $\lvert f\rvert \le \frac1{10}$ (for which $L=12$
gives full accuracy).

This completes the formulation for the area under a rhumb line for the
case of small flattening.  For larger eccentricities, we adopt the
technique introduced in \citet{karney-geod2}, namely to evaluate the
Fourier coefficients $P_l$ appearing in Eq.~(\ref{DeltaQ2}) at run time
by performing a discrete Fourier transform (DFT) on a set of evenly
spaced samples of $q_\beta(\beta)$.  The details of this approach are
given in Appendix \ref{area-app}.  Here, we merely note that a modest
number of Fourier coefficients allows the area to be computed with full
double-precision accuracy (see Table \ref{L-tab} in
Appendix~\ref{area-app}).

The term $p_{12}/\psi_{12}$ appearing in Eq.~(\ref{mean-xi}) presents
precisely the same issues as $\mu_{12}/\psi_{12}$ in Eq.~(\ref{s12}),
namely taking the ratio of two differences which is potentially
subject to an unacceptable loss of accuracy.  If $\phi_2 = \phi_1$, we
can use Eq.~(\ref{lhopital2}); but in the general case, we need
again to use divided differences as explained in
Appendix~\ref{area-app}.

It is interesting to compare this result, Eqs.~(\ref{mean-xi}) and
(\ref{Q-eq}), with the corresponding result for the geodesic problem,
Eqs.~(58), (59), and (61) in \citet{karney13}.  In both cases, we have
separated the area into a spherical contribution, which has singular
behavior for paths that pass close to a pole, and an ellipsoidal
correction.

However, in distinction to the geodesic problem, the longitude enters
the formula for the area, Eq.~(\ref{mean-xi}), as a multiplicative
factor, $\lambda_{12}$; this reflects the simple geometry of rhumb
lines.  As a consequence, $p_{12}$ depends just on the latitudes of the
endpoints and on the flattening of the ellipsoid.  This allows the
coefficients $P_l$ in Eq.~(\ref{DeltaQ2}) to be evaluated as soon as the
flattening is specified.  For the geodesic problem, this was not
possible, because the ellipsoidal correction, Eq.~(61)
in \citet{karney13}, involves both the flattening and the equatorial
azimuth of the geodesic inextricably linked together.

\section{Conclusions} \label{conclusions}

We have provided a method for computing the area under a segment of a
rhumb line.  The sum of the signed contributions for all the segments
making up a rhumb polygon gives the area of the polygon.  As with
geodesic polygons \citep{karney13}, a simple adjustment of the sum is
required if the polygon includes a pole.  This result for the area is
new.  \citet{freire10} have also considered the area of rhumb polygons,
but their approach involves breaking up the edges into sufficiently
short segments so that Simpson's method can be applied: the cost of the
area evaluation is then proportional to the perimeter of the polygon.
In contrast, the cost of the method described here is proportional to
the number of vertices in the polygon.  Remarkably, the comparatively
simple result for a sphere, when Eqs.~(\ref{Q-eq}) and (\ref{p0-eq})
reduce to $p(\chi) = \log\sec\chi$, is also new.

Key to maintaining accuracy in this work is the use of divided
differences.  Although these techniques are well
established \citep{kahan99}, they have been largely overlooked in
geodetic applications.  The application of divided differences to
Clenshaw summation is also new.

The methods described here were implemented as part of
GeographicLib \citep{geographiclib22} in 2014; however, the area
calculations were based on the Taylor expansion of $p_\chi(\chi)$ and
were therefore limited to small flattenings.  For this paper, the
routines were reimplemented to utilize the development of auxiliary
latitudes given in \citet{karney-auxlat}, to improve the treatment of
the area integral by the change of variables from $\beta$ to $\chi$, and
to compute the coefficients for the area integral using a DFT which
provided accurate results for the area for highly eccentric ellipsoids.
The interface consists of
\begin{itemize}
\item C++ classes {\tt Rhumb} and {\tt RhumbLine} to solve inverse
and direct rhumb line problems;
\item a command-line utility {\tt RhumbSolve}
which provides a convenient front end to these classes;
\item a command-line utility {\tt Planimeter}
to compute the area of geodesic and rhumb polygons;
\item online versions of these tools are also available
at \url{https://geographiclib.sourceforge.io/cgi-bin/RhumbSolve}
and \url{https://geographiclib.sourceforge.io/cgi-bin/Planimeter}.
\end{itemize}
This reimplementation is included in version 2.2 of GeographicLib.

\section*{
\ifeprint Supplementary data\else %%
{\it Supplementary data:}
\fi            %%
}
The following files are provided in
\begin{quote}
  \emph{Series expansions for computing rhumb areas},\\
  \doi{10.5281/zenodo.7685484},
\end{quote}
as supplementary data for this paper:
\begin{itemize}
\item {\tt rhumbarea.mac}:
the Maxima code used to obtain the series expansions for $\mathbf
P^{(L)}$, Eq.~(\ref{QdotN}).  Instructions for using this code are
included in the file.
\item {\tt auxlvals40.mac}:
The series for conversions between auxiliary latitudes which is required
by {\tt rhumbarea.mac}.  This is a copy of a file provided
in \citet{auxlatdata}.
\item {\tt rhumbvals40.mac}:
The matrix $\mathsf Q$ produced by {\tt rhumbarea.mac} for $L=40$ in a
form suitable for loading into Maxima.
\item {\tt rhumbvals\{6,16,40\}.m}:
The matrix $\mathsf Q$ in Octave/MATLAB notation for $L=6$ and $16$
(using exact fractions) and $L=40$ (written as floating-point numbers).
{\tt rhumbvals6.m} reproduces Eq.~(\ref{Qeq}) in
``machine-readable'' form.  {\tt rhumbvals40.m} can be used to estimate
the radii of convergence of Eq.~(\ref{QdotN}).  The format of these
files is sufficiently simple that they can be easily adapted for any
computer language.
\end{itemize}

\ifeprint\else                                           %%
\section*{{\it Funding:}}
No funds, grants, or other support was received.

\section*{{\it Disclosure statement:}}
The author has no conflict of interest.
\fi                                                      %%

\bibliography{geod}

\begin{thebibliography}{16}
\providecommand{\natexlab}[1]{#1}
\providecommand{\url}[1]{\texttt{#1}}
\providecommand{\urlprefix}{URL }
\expandafter\ifx\csname urlstyle\endcsname\relax
  \providecommand{\doi}[1]{doi:\discretionary{}{}{}#1}\else
  \providecommand{\doi}{doi:\discretionary{}{}{}\begingroup
  \urlstyle{rm}\Url}\fi
\providecommand{\eprint}[2][]{\url{#2}}

\bibitem[{Botnev and Ustinov(2014)}]{botnev14}
V.~A. Botnev and S.~M. Ustinov, 2014, \emph{Metody resheniya pryamoy i obratnoy
  geodezicheskikh zadach s vysokoy tochnost'yu ({M}ethods for direct and
  inverse geodesic problems solving with high precision)}, St. Petersburg State
  Polytechnical University Journal, \textbf{3}(198), 49--58,
  \urlprefix\url{https://ntv.spbstu.ru/fulltext/T3.198.2014_05.PDF}.

\bibitem[{Clenshaw(1955)}]{clenshaw55}
C.~W. Clenshaw, 1955, \emph{A note on the summation of {C}hebyshev series},
  Math. Comp., \textbf{9}(51), 118--120,
  \doi{10.1090/S0025-5718-1955-0071856-0}.

\bibitem[{Danielsen(1989)}]{danielsen89}
J.~S. Danielsen, 1989, \emph{The area under the geodesic}, Survey Review,
  \textbf{30}(232), 61--66, \doi{10.1179/003962689791474267}.

\bibitem[{Freire and de~Vasconcello(2010)}]{freire10}
R.~R. Freire and J.~C.~P. de~Vasconcello, 2010, \emph{Geodetic or rhumb line
  polygon area calculation over the {WGS-84} datum ellipsoid}, in \emph{XXIV
  FIG International Congress},
  \urlprefix\url{https://www.fig.net/resources/proceedings/fig_proceedings/fig2010/papers/fs02c/fs02c_freire_vasconcellos_3868.pdf}.

\bibitem[{Ivory(1798)}]{ivory98}
J.~Ivory, 1798, \emph{A new series for the rectification of the ellipsis},
  Trans. Roy. Soc. Edinburgh, \textbf{4}(2), 177--190,
  \doi{10.1017/S0080456800030817},
  \urlprefix\url{https://books.google.com/books?id=FaUaqZZYYPAC&pg=PA177}.

\bibitem[{Kahan and Fateman(1999)}]{kahan99}
W.~M. Kahan and R.~J. Fateman, 1999, \emph{Symbolic computation of divided
  differences}, SIGSAM Bull., \textbf{33}(2), 7--28,
  \doi{10.1145/334714.334716}.

\bibitem[{Karney(2013)}]{karney13}
C.~F.~F. Karney, 2013, \emph{Algorithms for geodesics}, J. Geodesy,
  \textbf{87}(1), 43--55, \doi{10.1007/s00190-012-0578-z}.

\bibitem[{Karney(2022{\natexlab{a}})}]{karney-geod2}
---, 2022{\natexlab{a}}, \emph{Geodesics on an arbitrary ellipsoid of
  revolution}, Technical report, SRI International, \eprint{2208.00492}.

\bibitem[{Karney(2022{\natexlab{b}})}]{karney-auxlat}
---, 2022{\natexlab{b}}, \emph{On auxiliary latitudes}, Technical report, SRI
  International, \eprint{2212.05818}.

\bibitem[{Karney(2022{\natexlab{c}})}]{auxlatdata}
---, 2022{\natexlab{c}}, \emph{Series expansions for converting between
  auxiliary latitudes}, \doi{10.5281/zenodo.7382666}.

\bibitem[{Karney(2023)}]{geographiclib22}
---, 2023, \emph{Geographic{L}ib, version 2.2},
  \urlprefix\url{https://geographiclib.sourceforge.io/C++/2.2}.

\bibitem[{Lambert(1772)}]{lambert72}
J.~H. Lambert, 1772, \emph{Notes and comments on the composition of terrestrial
  and celestial maps}, in R.~Caddeo and A.~Papadopoulos, editors,
  \emph{Mathematical Geography in the Eighteenth Century: Euler, Lagrange and
  Lambert (2022)}, pp. 367--422 (Springer),
  \doi{10.1007/978-3-031-09570-2\_16}, translated from German by A.
  A’Campo-Neuen,
  \urlprefix\url{https://books.google.com/books?id=o_s_MR3NUD4C}.

\bibitem[{Legendre(1811)}]{legendre11}
A.~M. Legendre, 1811, \emph{Exercices de Calcul Int\'egral sur Divers Ordres de
  Transcendantes et sur les Quadratures}, volume~1 (Courcier, Paris),
  \urlprefix\url{https://iris.univ-lille.fr/handle/1908/1541}.

\bibitem[{Maxima(2022)}]{maxima}
Maxima, 2022, \emph{A computer algebra system, version 5.46.0},
  \urlprefix\url{https://maxima.sourceforge.io}.

\bibitem[{Meyer and Rollins(2011)}]{meyer11}
T.~H. Meyer and C.~Rollins, 2011, \emph{The direct and indirect problem for
  loxodromes}, Navigation, \textbf{58}(1), 1--6,
  \doi{10.1002/j.2161-4296.2011.tb01787.x}.

\bibitem[{Olver \emph{et~al.}(2010)Olver, Lozier, Boisvert, and Clark}]{dlmf10}
F.~W.~J. Olver, D.~W. Lozier, R.~F. Boisvert, and C.~W. Clark, editors, 2010,
  \emph{{NIST} Handbook of Mathematical Functions} (Cambridge Univ. Press),
  \urlprefix\url{https://dlmf.nist.gov}.

\end{thebibliography}

\appendix

\section{Use of divided differences} \label{divided}

Here I explore how to compute expressions such as $\mu_{12}/\psi_{12}$,
which appears in Eqs.~(\ref{s12}) and (\ref{lam12}), in a way that
maintains accuracy.  Such expressions can be treated in a systematic way
using the algebra of {\it divided differences}; the subject is discussed
in detail by \citet[henceforth referred to as {\it KF}]{kahan99}.  The
divided difference of $f(x)$ is defined by
\begin{equation}\label{divdiff}
 \divdiff[f](x,y) =
\begin{cases}
df(x)/dx, & \text{if $x=y$,} \\[0.5ex]
\displaystyle\frac{f(y)-f(x)}{y-x}, & \text{otherwise.}
\end{cases}
\end{equation}
The second form above can be used in a computation provided that $x$ and
$y$ are sufficiently far apart.  The key is to recast this expression
into one that avoids excessive roundoff error when $y \approx x$.

Many readers will already be familiar with suitable substitutions for
some cases, e.g.,
\begin{equation}
\divdiff[\sin](x,y) = {\textstyle \cos\frac12(x+y)} \frac{\sin\frac12(y-x)}
{\frac12(y-x)},
\end{equation}
which can be evaluated accurately if $y\approx x$ even if the {\it
relative} error in $y-x$ is significant; because the last factor above
varies slowly near $y-x = 0$, this error in the divided difference
depends only on the {\it absolute} error in $y-x$.  Section 1.4 in {\it
KF} provides a set of rules which allow the technique to be applied to a
wide range of problems.  In particular, suitable divided difference
formulas are given for all the elementary functions.

If we regard $\mu$ and $\psi$ as functions of $\chi$, we can write
\begin{equation}\label{s12-divdiff}
\frac{\mu_{12}}{\psi_{12}}
= \frac{\divdiff[\mu](\chi_1, \chi_2)} {\divdiff[\psi](\chi_1, \chi_2)}.
\end{equation}
In the spherical limit, where $\mu = \chi$, the numerator reduces to
unity.  Therefore we start by considering the denominator, which can be
expressed as
\ifeprint                                             %%
\begin{multline}                                      %%
\label{divdiff-psi}                                   %%
\divdiff[\psi](\chi_1, \chi_2)                        %%
=\divdiff[\sinh^{-1}](\tan\chi_1, \tan\chi_2)\\       %%
\times\divdiff[\tan](\chi_1, \chi_2),                 %%
\end{multline}                                        %%
\else                                                 %%
\begin{equation}
\label{divdiff-psi}
\divdiff[\psi](\chi_1, \chi_2)
=\divdiff[\sinh^{-1}](\tan\chi_1, \tan\chi_2)
\divdiff[\tan](\chi_1, \chi_2),
\end{equation}
\fi                                                   %%
where I have applied the chain rule
\begin{equation}\label{div-chain}
\divdiff[f\circ g](x,y) =
\divdiff[f]\bigl(g(x),g(y)\bigr)\,\divdiff[g](x,y)
\end{equation}
(here $f\circ g$ refers to functional composition) to
Eq.~(\ref{psi-chi}).  Using $\tan\chi$ as the independent variable
(instead of $\chi$) allows us to represent accurately values of $\chi$
close to $0$ and $\pm\frac12\pi$; thus we use the inverse function rule,
\begin{equation}\label{div-inv}
\divdiff[f^{-1}](x,y) =1/\divdiff[f]\bigl(f^{-1}(x),f^{-1}(y)\bigr),
\end{equation}
to rewrite Eq.~(\ref{divdiff-psi}) as
\begin{equation}\label{divdiff-psi2}
\divdiff[\psi](\chi_1, \chi_2)
=\frac{\divdiff[\sinh^{-1}](\tan\chi_1,\tan\chi_2)}
  {\divdiff[\tan^{-1}](\tan\chi_1,\tan\chi_2)}.
\end{equation}
Note that the chain rule and inverse function rules,
Eqs.~(\ref{div-chain}) and (\ref{div-inv}), are direct analogs of the
corresponding rules for derivatives.  Equation (\ref{divdiff-psi2}) can
be evaluated using the formulas from {\it KF}:
\begin{align}
\divdiff[\sinh^{-1}](x,y) &=
\sinh^{-1}\biggl(\frac{(y-x)(x+y)}{x\sqrt{1+y^2}+y\sqrt{1+x^2}}\biggr)\notag\\
&\qquad{}\times \frac1{y-x},\label{asinh-divdiff}
\displaybreak[0]\\
\divdiff[\tan^{-1}](x,y) &= \frac1{y-x}\tan^{-1}\frac{y - x}{1 + xy}.
\label{atan-divdiff}
\end{align}
For a sphere, where $\divdiff[\mu](\chi_1, \chi_2) = 1$, Eq.~(\ref{s12})
can now be evaluated using Eqs.~(\ref{s12-divdiff}) and
(\ref{divdiff-psi2}).  Repeating the numerical experiment of
Fig.~(\ref{roundoff}), the roundoff errors using divided differences are
shown as diamonds; this confirms that this method is effective in
dramatically reducing the numerical error.

I've only included the basic results in Eqs.~(\ref{asinh-divdiff}) and
(\ref{atan-divdiff}); {\it KF} supplement these with the derivatives if
$y-x$ vanishes or the direct ratios if $x$ and $y$ have opposite signs.
Also, it is wise to ensure that correct results are returned in ``corner
cases'', e.g., $x = 0$ and $y = \infty$.

Turning now to the numerator of Eq.~(\ref{s12-divdiff}),
$\divdiff[\mu](\chi_1, \chi_2)$.  We distinguish two cases: if the
flattening of the ellipsoid is small, say $\lvert
f\rvert \le \frac1{100}$, then we can use a sixth-order trigonometric
series to convert between $\chi$ and $\mu$.  This can be evaluated using
Clenshaw summation and to compute the divided difference we need to
extend Clenshaw summation.  Since this method is generally useful, we
address this separately in Appendix~\ref{clenshaw}.

For arbitrary flattening, we recast Eq.~(\ref{s12-divdiff}) as
\begin{equation}\label{s12-divdiff2}
\frac{\mu_{12}}{\psi_{12}} =
\frac{\divdiff[\mu](\beta_1, \beta_2)\,\divdiff[\beta](\phi_1, \phi_2)}
{\divdiff[\psi](\phi_1, \phi_2)},
\end{equation}
and apply divided differences to
\begin{align}
\beta(\phi) &= \tan^{-1}\bigl((1-f)\tan\phi\bigr),\label{beta-phi}
\displaybreak[0]\\
\mu(\beta) &= \frac{bE(\beta, ie')}R,\label{mu-beta}
\displaybreak[0]\\
\psi(\phi) &= \lam\phi - e\tanh^{-1}(e\sin\phi),\label{psi-phi}
\end{align}
where $E(x, k)$ is the incomplete elliptic integral of the second
kind \citep[\S\dlmf{19.2(ii)}{19.2.ii}]{dlmf10}.  For
Eqs.~(\ref{beta-phi}) and (\ref{psi-phi}), this is just a routine
exercise applying the results of {\it KF}.

{\it KF} do not provide a formula for $\divdiff[E](x,y;k)$; they do
however note ``Examination of `addition formulas' may provide some
suggestions for divided differences of additional special functions.''
Indeed, the necessary addition formula is found
in \citet[Eq.~\dlmf{19.11.2}{19.11.E2}]{dlmf10}.  With the substitutions
$\theta = y$, $\phi = -x$, $\psi = z$, this can be converted to the
following divided difference formula,
\begin{equation}
\divdiff[E](x,y;k)
=\begin{cases}
\sqrt{1 - k^2\sin^2x}, & \text{if $x=y$,} \\[0.5ex]
\displaystyle \frac{E(y,k)-E(x,k)}{y-x}, & \text{if $xy \le0$,}\\[0.5ex]
\displaystyle
\biggl(\frac{E(z,k)}{\sin z} - k^2 \sin x \sin y\biggr)\\
\qquad\displaystyle{}\times\frac{\sin z}{y-x},
&\text{otherwise,}
\end{cases}
\end{equation}
where the angle $z$ is given by
\begin{align}
z &= \atanx{2t}{(1-t)(1+t)},\displaybreak[0]\\
t &= \frac{(y-x)\divdiff[\sin](x,y)}
{\sin x\sqrt{1 - k^2\sin^2y} + \sin y\sqrt{1 - k^2\sin^2x}}\notag\\
&\qquad{}\times\frac{\sin x + \sin y}{\cos x + \cos y}.
\end{align}
(Note that the quantity $t$ is just $\tan\frac12z$.)  This formula is
restricted to the case $\lvert x\rvert \le \frac12\pi$ and $\lvert
y\rvert \le \frac12\pi$.  It is best used for $k^2 < 0$ so that the two
terms in large parentheses have the same sign; this is the case for an
oblate ellipsoid ($e'^2 > 0$), since $k^2 = -e'^2 < 0$.  For a prolate
ellipsoid ($e^2 < 0$), we can apply finite differences to $\mu'(\beta')$
(the primes indicate the complementary angles), where
\begin{align}
\divdiff[\mu](\beta_1, \beta_2) &= \divdiff[\mu'](\beta'_1, \beta'_2),
\displaybreak[0]\\
\mu'(\beta') &= \frac{aE(\beta', e)}R,\label{mu1-beta}
\end{align}
and once again we have $k^2 = e^2 < 0$.

We have given here a cursory examination of using divided differences
for the ellipsoid with arbitrary flattening.  Full details are available
in the implementation described in Sec.~\ref{conclusions}.

\section{Divided differences for Clenshaw summation}
\label{clenshaw}

The numerator of Eq.~(\ref{s12-divdiff}) is
$\divdiff[\mu](\chi_1, \chi_2)$.  Here, we examine evaluating
$\divdiff[\eta](\zeta_1, \zeta_2)$ for any two auxiliary latitudes
$\zeta$ and $\eta$ in the limit of small flattening.  In this case, we
can approximate \citep{karney-auxlat}
\begin{align}
\eta(\zeta) &= \zeta + \Delta \eta(\zeta) + O(n^{L+1}), \label{aux-series}
\displaybreak[0]\\
\Delta \eta(\zeta) &= \sum_{l=1}^L C_l \sin 2l\zeta,
\label{aux-clen}
\end{align}
where choosing $L=6$ suffices to provide accurate results for $\lvert
f\rvert \le \frac1{150}$.  The additive property of divided differences
lets us write
\begin{equation}\label{aux-divdiff}
\divdiff[\eta](\zeta_1, \zeta_2) = 1 +
\divdiff[\Delta\eta](\zeta_1, \zeta_2).
\end{equation}
Because in computing the area under a rhumb line we also need to compute
a cosine sum, we consider the more general sum
\begin{equation}\label{trigsum}
p(\zeta) = \sum_{k=0}^{K-1} c_k f_k(\zeta),
\end{equation}
where
\begin{equation}\label{fk}
f_k(\zeta) = \sin\bigl(2(k+k_0)\zeta + \zeta_0\bigr).
\end{equation}
For the conversions between auxiliary latitudes, Eq.~(\ref{aux-clen}),
we have $k_0=1$ and $\zeta_0=0$, $f_k(\zeta) = \sin2{(k+1)}\zeta$;
for computing areas, Eq.~(\ref{DeltaQ2}), we have $k_0=1$ and
$\zeta_0=\frac12\pi$, $f_k(\zeta) = \cos2{(k+1)}\zeta$.

\citet{clenshaw55} summation can be applied to
Eqs.~(\ref{trigsum}) and (\ref{fk}) as follows: Noting that $f_k(\zeta)$
obeys the recurrence relation
\begin{equation}\label{recur1}
f_{k-1}(\zeta) + f_{k+1}(\zeta) = 2x f_k(\zeta),
\end{equation}
where
\begin{equation}\label{recur2}
x = \cos2\zeta = (\cos\zeta + \sin\zeta)(\cos\zeta - \sin\zeta),
\end{equation}
we have
\begin{align}
  u_k &= \begin{cases} 0, &\text{for $k \ge K$,}\\
 2 x u_{k+1} - u_{k+2} + c_k, &\text{for $k < K$},
 \end{cases}
  \label{clen2}\displaybreak[0]\\
  p(\zeta) &= u_0 f_0(\zeta) - u_1 f_{-1}(\zeta). \label{clen3}
\end{align}
This method of summation has the advantages: (1)~minimal evaluation of
trigonometric functions, (2)~the code executes a tight loop as it reads
an array of coefficients, and (3)~the terms are summed in reverse order
(which in the usual case implies from smallest to largest).

Applying divided differences, Eq.~(\ref{divdiff}), to
Eq.~(\ref{trigsum}) gives
\begin{equation}
\divdiff[p](\zeta_1, \zeta_2) =
\sum_{k=0}^{K-1} c_k \divdiff[f_k](\zeta_1, \zeta_2),
\end{equation}
with
\begin{equation}
\divdiff[f_k](\zeta_1, \zeta_2)
=2 \cos\bigl((k+k_0)\zeta_+ + \zeta_0\bigr)
\frac{\sin(k+k_0)\zeta_-}{\zeta_-},
\end{equation}
where $\zeta_- = \zeta_2-\zeta_1$, $\zeta_+ = \zeta_2+\zeta_1$, and in
the limit $\zeta_- \rightarrow 0$, we have
$\bigl(\sin{(k+k_0)}\zeta_-\bigr)/\zeta_-\allowbreak\rightarrow k +
k_0$.  While this form avoids the loss of accuracy when
$\zeta_2 \approx \zeta_1$, we have lost the first two advantages listed
above for Clenshaw summation.

Here we give a generalization of Clenshaw summation which allows for the
sum and difference of $p(\zeta_{1,2})$ to be computed together.  We
write
\begin{equation}
\mathbf P(\zeta_1,\zeta_2) = \begin{bmatrix}
p(\zeta_2) + p(\zeta_1)\\[0.5ex]
\displaystyle\frac{p(\zeta_2) - p(\zeta_1)}\Delta
\end{bmatrix} = \sum_{k=0}^{K-1} \mathbf F_k(\zeta_1,\zeta_2),
\end{equation}
where
\begin{equation}
\mathbf F_k(\zeta_1,\zeta_2) = \begin{bmatrix}
f_k(\zeta_2) + f_k(\zeta_1)\\[0.5ex]
\displaystyle\frac{f_k(\zeta_2) - f_k(\zeta_1)}\Delta
\end{bmatrix},
\end{equation}
and we consider either $\Delta = 1$, where the second component of
$\mathbf P(\zeta_1,\zeta_2)$ is the {\it plain} difference, or $\Delta
= \zeta_2-\zeta_1$, where the second component is the {\it divided}
difference, $\divdiff[p](\zeta_1, \zeta_2)$.

Equations (\ref{recur1}) and (\ref{recur2}) generalize to
\begin{equation}
\mathbf F_{k-1}(\zeta_1,\zeta_2) + \mathbf F_{k+1}(\zeta_1,\zeta_2) =
2\mathsf X \cdot \mathbf F_k(\zeta_1,\zeta_2),
\end{equation}
where
\begin{equation}
\mathsf X =
\begin{bmatrix}
\cos\zeta_- \cos\zeta_+
&-\Delta\sin\zeta_- \sin\zeta_+
\\[0.5ex]
-\displaystyle\frac{\sin\zeta_-}\Delta \sin\zeta_+
&\cos\zeta_- \cos\zeta_+
\end{bmatrix}.
\end{equation}
Finally, Eqs.~(\ref{clen2}) and (\ref{clen3}) are replaced by
\begin{align}
\mathsf U_k &= \begin{cases} \mathsf 0,& \text{for $k \ge K$},\\
2\mathsf X \cdot \mathsf U_{k+1} - \mathsf U_{k+2}  +
c_k \mathsf I,& \text{for $k < K$},
\end{cases}\displaybreak[0]\\
\mathbf P(\zeta_1,\zeta_2) &= \mathsf U_0\cdot \mathbf F_0(\zeta_1,\zeta_2)
-\mathsf U_1\cdot \mathbf F_{-1}(\zeta_1,\zeta_2).
\end{align}
For evaluating the series for auxiliary latitudes as in
Eq.~(\ref{aux-clen}), we set $f_k(\zeta) = \sin2{(k+1)}\zeta$ and we
obtain
\begin{equation}
\mathbf F_{-1}(\zeta_1,\zeta_2) = \mathbf 0
, \quad
\mathbf F_0(\zeta_1,\zeta_2) = 2\begin{bmatrix}
\sin\zeta_+ \cos\zeta_-\\[0.5ex]
\cos\zeta_+ \displaystyle\frac{\sin\zeta_-}\Delta
\end{bmatrix}.
\end{equation}
Likewise, for the area series Eq.~(\ref{DeltaQ2}), we set $f_k(\zeta)
= \cos2{(k+1)}\zeta$ which yields
\begin{equation}
\mathbf F_{-1}(\zeta_1,\zeta_2) = \begin{bmatrix}2\\0\end{bmatrix}
, \quad
\mathbf F_0(\zeta_1,\zeta_2) = 2\begin{bmatrix}
\cos\zeta_+ \cos\zeta_-\\[0.5ex]
-\sin\zeta_+ \displaystyle\frac{\sin\zeta_-}\Delta
\end{bmatrix}.
\end{equation}
The $2\times2$ matrices, $\mathsf X$ and $\mathsf U_k$, are of the form
\begin{equation}
\begin{bmatrix}
x&\Delta^2y\\
y&x
\end{bmatrix}.
\end{equation}
Multiplication and addition of such matrices yield matrices of the same
form.  In carrying out the matrix operations computationally, it is
therefore possible to compute and store just the first column of the
matrix.  The cost of computing the divided difference,
$\divdiff[p](\zeta_1, \zeta_2)$, in this more accurate fashion, is then
only modestly more than the cost of computing $p(\zeta_1)$ and
$p(\zeta_2)$ separately.

\section{Evaluating the area integral}
\label{area-app}

Here we cover two issues with computing the area $S_{12}$ from
Eq.~(\ref{mean-xi}): the use of divided differences to
$p_{12}/\psi_{12}$, and evaluating $P_l$ appearing in
Eq.~(\ref{DeltaQ2}) when the eccentricity is large.

Addressing the first issue, we apply divided differences to
$p(\chi) = p_0(\chi) + p_\beta\bigl(\beta(\chi)\bigr)$ to obtain
\ifeprint                                                               %%
\begin{multline}\label{area-divdiff}                                    %%
\frac{p_{12}}{\psi_{12}}                                                %%
= \frac{\divdiff[p_0](\chi_1, \chi_2)}                                  %%
{\divdiff[\psi](\chi_1, \chi_2)}\\                                      %%
{}+\divdiff[p_\beta](\beta_1, \beta_2)                                  %%
\frac{\divdiff[\beta](\chi_1,\chi_2)}{\divdiff[\psi](\chi_1, \chi_2)}.  %%
\end{multline}                                                         %%
\else                                                                  %%
\begin{equation}\label{area-divdiff}
\frac{p_{12}}{\psi_{12}}
= \frac{\divdiff[p_0](\chi_1, \chi_2)}
{\divdiff[\psi](\chi_1, \chi_2)}
+\divdiff[p_\beta](\beta_1, \beta_2)
\frac{\divdiff[\beta](\chi_1,\chi_2)}{\divdiff[\psi](\chi_1, \chi_2)}.
\end{equation}
\fi                                                                    %%
For the first (spherical) term in Eq.~(\ref{area-divdiff}), we
rewrite Eq.~(\ref{p0-eq}) as
\begin{align}
p_0(\chi) &= \sinh^{-1} h(\tan\chi),\displaybreak[0]\\
h(x) &= \frac{x^2}{2\sqrt{1+x^2}}.
\end{align}
We now have
\ifeprint                                                       %%
\begin{multline}                                                %%
\frac{\divdiff[p_0](\chi_1, \chi_2)}                            %%
{\divdiff[\psi](\chi_1, \chi_2)} = \frac                        %%
{\divdiff[\sinh^{-1}]\bigl(h(\tan\chi_1), h(\tan\chi_2)\bigr)}  %%
{\divdiff[\sinh^{-1}](\tan\chi_1, \tan\chi_2)}\\[0.5ex]         %%
{}\times\divdiff[h](\tan\chi_1, \tan\chi_2),                    %%
\end{multline}                                                  %%
\else                                                           %%
\begin{equation}
\frac{\divdiff[p_0](\chi_1, \chi_2)}
{\divdiff[\psi](\chi_1, \chi_2)} = \frac
{\divdiff[\sinh^{-1}]\bigl(h(\tan\chi_1), h(\tan\chi_2)\bigr)}
{\divdiff[\sinh^{-1}](\tan\chi_1, \tan\chi_2)}
\divdiff[h](\tan\chi_1, \tan\chi_2),
\end{equation}
\fi                                                             %%
where straightforward algebraic simplification gives
\ifeprint                                     %%
\begin{multline}                              %%
\divdiff[h](x, y) = \\                        %%
\frac{(x+y) (x^2 + y^2 + x^2y^2)}             %%
{2\sqrt{1+x^2}\sqrt{1+y^2}                    %%
\bigl(x^2\sqrt{1+y^2}+y^2\sqrt{1+x^2}\bigr)}. %%
\end{multline}                                %%
\else                                         %%
\begin{equation}
\divdiff[h](x, y) =
\frac{(x+y) (x^2 + y^2 + x^2y^2)}
{2\sqrt{1+x^2}\sqrt{1+y^2}
\bigl(x^2\sqrt{1+y^2}+y^2\sqrt{1+x^2}\bigr)}.
\end{equation}
\fi                                           %%

Turning now to the second (ellipsoidal) term in
Eq.~(\ref{area-divdiff}), the first factor,
$\divdiff[p_\beta](\beta_1, \beta_2)$, can be straightforwardly
evaluated using the divided difference method for Clenshaw sums (see
Appendix \ref{clenshaw}) because we represent $p_\beta(\beta)$ as a
cosine series, Eq.~(\ref{DeltaQ2}); this applies to our approaches both
for small flattening and for arbitrary eccentricity.

For the second factor in the ellipsoidal term in
Eq.~(\ref{area-divdiff}),
$\divdiff[\beta](\chi_1,\chi_2)/\divdiff[\psi](\chi_1, \chi_2)$, we
again distinguish two cases.  If the flattening is small, we can
represent $\beta(\chi)$ as a series of the form given in
Eqs.~(\ref{aux-series}) and (\ref{aux-clen}), and this term can be
evaluated using Eqs.~(\ref{divdiff-psi2}) and (\ref{aux-divdiff}).  If
the flattening is large, both $\beta$ and $\psi$ should be treated as
functions of $\phi$ giving
\begin{equation}
\frac{\divdiff[\beta](\chi_1,\chi_2)}{\divdiff[\psi](\chi_1, \chi_2)}
=\frac{\divdiff[\beta](\phi_1,\phi_2)}{\divdiff[\psi](\phi_1, \phi_2)},
\end{equation}
which can be evaluated by applying divided differences to
Eqs.~(\ref{beta-phi}) and (\ref{psi-phi}).

The remaining task is to prescribe how $P_l$ in Eq.~(\ref{DeltaQ2}) can
be evaluated for large eccentricity.  \citet{karney-geod2} establishes
that these coefficients can be computed by performing a DFT on equally
sampled values of the integrand $q_\beta(\beta)$, Eq.~(\ref{q-eq-beta});
this provides an accurate method for numerically computing the
indefinite integral of a periodic function.

Naturally, when evaluating $q_\beta(\beta)$ numerically we should replace
Eq.~(\ref{q-eq-beta}) by
\begin{equation}\label{q-eq-beta-alt}
q_\beta(\beta)
= (1-f) \frac{\cos\chi-\cos\xi}{\cos\phi}
\frac{\cos\xi+\cos\chi}{\sin\xi+\sin\chi},
\end{equation}
when $\lvert\sin\chi\rvert > \lvert\cos\chi\rvert$, in order to minimize
the roundoff error.  Here $\chi$ and $\xi$ are considered as functions
of $\beta$ and these functions must be implemented in such a way that
the {\it relative} errors in $\cos\chi$ and $\cos\xi$ are small; the
conversion formulas that are given in \citet{karney-auxlat} fulfill this
condition.

The function $q_\beta(\beta)$ is an odd periodic function of $\beta$
with period $\pi$ and so can be written as
\begin{equation}\label{qbetasum}
q_\beta(\beta) = \sum_{l=1}^\infty b_l \sin2l\beta.
\end{equation}
However, because the routines used to compute the DFT of the geodesic
area integral in \citet{karney-geod2} were specialized to DST-III and
DST-IV transforms, we elect to work with
\begin{equation}
\frac{q_\beta(\beta)}{\cos\beta} = \sum_{l=1}^\infty c_l \sin(2l-1)\beta,
\end{equation}
where
\begin{equation}\label{coeff-int}
c_l = \frac4\pi \int_0^{\pi/2} \frac{q_\beta(\beta)}{\cos\beta}
\sin(2l-1)\beta\,\d\beta.
\end{equation}
Now the Fourier coefficients for $q_\beta(\beta)$ and $p_\beta(\beta)$
in Eqs.~(\ref{qbetasum}) and (\ref{DeltaQ2}) are
\begin{align}
b_l &= \frac{c_l+c_{l+1}}2,\displaybreak[0]\\
P_l &= -\frac{c_l+c_{l+1}}{4l}.\label{P-c}
\end{align}

\begin{table}[tbp]
\caption{\label{L-tab}
% Columns L + L' are cols 2+3 from BUILD5/areaint
% Column L_\chi is col 2 from BUILD5/areaint with CHI_AREA_SERIES=1
The required number of terms in the Fourier series for the ellipsoidal
contribution to the area $p_\beta(\beta)$, Eq.~(\ref{DeltaQ2}), as a
function of the third flattening $n$.  $L$ is the minimum number of
coefficients to provide full double-precision accuracy.  $L'$ is the
estimate of the number of coefficients using a simple rule that can be
implemented in double precision.  $L_\chi$ is the minimum number of
coefficients that are required for accurately representing
$q_\chi(\chi)$; here the notation ``$-$'' indicates that $L_\chi >
2^{14}$.}
\begin{center}
\newcommand{\0}{\hphantom{0}}
\newcommand{\1}{\hphantom{\pm}}
\begin{tabular}{@{\extracolsep{0.3em}}
>{\rule{0.7em}{0pt}$}c<{$}>{$}r<{$}>{$}r<{$}>{$}r<{\rule{1.2em}{0pt}$}|
>{\rule{0.4em}{0pt}$}c<{$}>{$}r<{$}>{$}r<{$}>{$}r<{\rule{0.7em}{0pt}$}}
\hline\hline
      n &    L &   L' &L_\chi &     n &    L &    L' &L_\chi
    \rule[-1.5ex]{0pt}{4.5ex}\\\hline
\pm0.01 &    6 &    7 &     7 \rule[0ex]{0pt}{3ex}\\
\pm0.02 &    7 &    8 &     9 \\
\1 0.05 &    9 &   11 &    12 & -0.05 &    8 &   10 &    13 \\
\1 0.10 &   11 &   13 &    17 & -0.10 &   12 &   14 &    18 \\
\1 0.20 &   16 &   19 &    28 & -0.20 &   16 &   19 &    35 \\
\1 0.30 &   21 &   24 &    44 & -0.30 &   21 &   24 &    69 \\
\1 0.40 &   27 &   30 &    68 & -0.40 &   27 &   31 &   158 \\
\1 0.50 &   34 &   38 &   107 & -0.50 &   35 &   39 &   473 \\
\1 0.60 &   44 &   50 &   179 & -0.60 &   46 &   51 &  2352 \\
\1 0.70 &   60 &   66 &   326 & -0.70 &   63 &   69 &   -\0 \\
\1 0.80 &   89 &   95 &   712 & -0.80 &   97 &  108 &   -\0 \\
\1 0.90 &  164 &  167 &  2383 & -0.90 &  205 &  222 &   -\0 \\
\1 0.95 &  288 &  276 &  7096 & -0.95 &  421 &  446 &   -\0 \\
\1 0.99 &\0942 &\0688 &   -\0 & -0.99 & 2146 & 2163 &   -\0
\rule[-1.5ex]{0pt}{1.5ex}\\
\hline\hline
\end{tabular}
\end{center}
\end{table}
In order to evaluate the integral Eq.~(\ref{coeff-int}) for $c_l$, we
sample $q_\beta(\beta)/\cos\beta$ at $L$ equally spaced points over a
quarter period and perform a discrete sine transform on these values.
This gives an estimate of the first $L$ coefficients $c_l$ and, from
Eq.~(\ref{P-c}), $P_l$.  We are confronted with the question of picking
a suitable value of $L$ to guarantee that the absolute error in
$p_\beta(\beta)$ is no more than $2^{-53}$ (the roundoff limit for
double precision).  This value of $L$ can be readily found using high
(256-bit) precision arithmetic; the results are shown in the column
labeled $L$ in Table \ref{L-tab} for selected values of the third
flattening $n$.  These results are surprisingly close to the
corresponding values, $N$, for the geodesic area
problem \citep[Fig.~5]{karney-geod2}.

We would like to be able to estimate $L$ at run time when high-precision
arithmetic is not available.  In this case, computing the error is
confounded by roundoff.  So instead we determine $L'$, the minimum value
of $L$ such that a fraction $\frac18$ of the last coefficients $P_l$ are
all less than $2^{-53}$ in magnitude.  The resulting values of $L'$ are
also given in Table \ref{L-tab}.  This approximation to $L$ is slightly
conservative ($L'$ modestly bigger than $L$) for $n\le 0.92$ or $a/b \le
24$, i.e., for all but highly oblate ellipsoids.

Determining $L'$ can be done inexpensively because increasing the number
of samples used to compute the DFT from $L$ to $2L$ can be carried out
by combining the order $L$ Fourier coefficients with a separate order
$L$ calculation using interleaved samples.  This allows $P_l$ to be
computed once the flattening of the ellipsoid is specified.

The column headed $L_\chi$ in Table \ref{L-tab} shows the number of
Fourier coefficients we would have required had we computed the Fourier
series for $q_\chi(\chi)$.  The larger values of $L_\chi$ reflect the
near singular behavior of $q_\chi(\chi)$ exhibited in
Fig.~\ref{qintegrand} and this shows the dramatic benefit of making the
change of variables from $\chi$ to $\beta$.

It is instructive to compare the two approaches for approximating $P_l$
for a given $L$: using a Taylor series truncated at order $n^L$ or using
the DFT on $L$ samples of $q_\beta(\beta)$.
\begin{itemize}
\item
The error in the individual coefficients is smaller using the DFT method.
\item
Using the Taylor series method requires storing the upper triangular
matrix $\mathsf Q^{(L)}$, i.e., about $\frac12L^2$ numbers.  In
contrast, the DFT method requires only storing coefficients $P_l$, i.e.,
$L$ values.
\item
The DFT requires $L$ evaluations of $q_\beta(\beta)$ and a subsequent
DFT to compute $c_l$ and $P_l$ with cost $O(L\log L)$.  Evaluating $P_l$
with the Taylor series method requires the matrix-vector product,
Eq.~(\ref{QdotN}), at a cost of $\frac12L^2$; of course, each of these
operations is very simple compared, for example, to the cost of
computing a single value of $q_\beta(\beta)$.
\item
The Taylor series method entails very simple programming, while the DFT
method requires the complex machinery for fast Fourier transforms.
\end{itemize}
These considerations reinforce the conclusion that the approach using
Taylor series is appropriate for small flattening where a small value
$L$ is adequate, while the DFT method is preferred for larger
eccentricities.

We close by giving a numerical example of the area computation: consider
an ellipsoid with $a = 6400\,\mathrm{km}$ and $f = \frac15$ and a
rhumb segment starting on the equator at $\phi_1=0^\circ$,
$\lambda_1 = 0^\circ $, with azimuth $\alpha_{12} = 45^\circ$ and length
$s_{12} = 2000\,\mathrm{km}$.  The endpoint of this segment $\phi_2 =
19.380\,181\,12^\circ$, $\lambda_2 = 12.823\,427\,61^\circ$ and the area
between it and the equator is $S_{12} =
1\,012\,834.108\,565\,\mathrm{km}^2$.  Readers can generate additional
examples using the online {\tt RhumbSolve} tool mentioned at the end of
Sec.~\ref{conclusions}.

\end{document}